\documentclass[pre,amsmath,amssymb,twocolumn,superscriptaddress]{revtex4-2}
\usepackage{graphicx}
\usepackage{xcolor}
\usepackage{braket}
\usepackage{bbold}
\usepackage{hyperref}
\hypersetup{
        citecolor  = blue,
        colorlinks = true,
    }
\usepackage{siunitx}
\usepackage{natbib}

\newcommand{\tr}{{\rm tr}}

\newcommand{\im}{{\rm i}}

\newcommand{\comment}[1]{}

\newcommand{\sutdphys}{Science, Mathematics and Technology Cluster, Singapore
University of Technology and Design, 8 Somapah Road, 487372 Singapore}
\newcommand{\sutdepd}{EPD Pillar, Singapore University of Technology and Design, 8 Somapah Road, 487372 Singapore}
\newcommand{\entropica}{Entropica Labs, 186b Telok Ayer Street, 068632 Singapore}
\newcommand{\cqt}{Centre for Quantum Technologies, National University of Singapore 117543, Singapore} 
\newcommand{\majulab}{MajuLab, CNRS-UNS-NUS-NTU International Joint Research Unit, UMI 3654, Singapore}

\begin{document}
\title{Simulating quantum transport via collisional models on a digital quantum computer}

\author{Rebecca Erbanni}
\affiliation{\sutdphys}

\author{Xiansong Xu} 
\affiliation{\sutdphys}

\author{Tommaso Demarie}
\affiliation{\entropica}    

\author{Dario Poletti} 
\affiliation{\sutdphys}
\affiliation{\sutdepd}
\affiliation{\cqt} 
\affiliation{\majulab} 
\begin{abstract}
Digital quantum computers have the potential to study the dynamics of complex quantum systems. Nonequilibrium open quantum systems are, however, less straightforward to be implemented. Here we consider a collisional model representation of the nonequilibrium open dynamics for a boundary-driven XXZ spin chain, with a particular focus on its steady states. More specifically, we study the interplay between the accuracy of the result versus the depth of the circuit by comparing the results generated by the corresponding master equations. We study the simulation of a boundary-driven spin chain in regimes of weak and strong interactions, which would lead in large systems to diffusive and ballistic dynamics, considering also possible errors in the implementation of the protocol. Last, we analyze the effectiveness of digital simulation via the collisional model of current rectification when the XXZ spin chains are subject to non-uniform magnetic fields. 
\end{abstract}
\maketitle 

\section{Introduction} 
The study of quantum transport in boundary-driven many-body quantum systems is a significant theoretical, numerical, and experimental challenge \cite{BertiniZnidaric2021, LandiSchaller2022}.  Additional insights from quantum simulators and digital quantum computers can help us to further explore and understand the phenomenology of these systems. However, while these platforms have been used to study isolated quantum systems extensively, the study of open quantum dynamics with controllable environment settings is still an ongoing challenge. A possible way to model these systems is via collisional models  \cite{KarplusSchwinger1948,RauRau1963,DumckeDumcke1985, Burgarth_2007, CiccarelloPalma2022, CampbellVacchini2021, CattaneoManiscalco2022, CattaneoGiorgi2021}, also known as repeated interactions schemes. 
Recently these collisional models have attracted significant attention from the field of quantum thermodynamics \cite{ScaraniBuzek2002, StrasbergEsposito2017a}. For instance, they have been used to study nonequilibrium steady states \cite{KarevskiPlatini2009a,SeahScarani2019a, GuarnieriCampbell2020}, charging of batteries \cite{SeahNimmrichter2021}, thermometry \cite{SeahLandi2019}, non-thermal baths \cite{ShuScarani2019}, and current rectification \cite{LandiKarevski2014a}. Collisional models have also been proven useful to maintain a framework of energy transport consistent with the laws of thermodynamics \cite{Pereira2018, DeChiaraAntezza2018, HoferBrunner2017} and, to further show their versatility, have recently been used to model a perceptron in the context of binary classification of quantum data \cite{Korkmaz_2023}.   

In general, such a framework treats the environment as a set of ancillae with which the system interacts sequentially for a finite time $\tau$. Hence, the dynamics of the system and ancillae can be described by a sequence of unitary operations that can be more readily implemented on quantum computers and other experimental platforms \cite{CuevasMataloni2019, Garcia-PerezManiscalco2020,BurgerPoletti2022, CattaneoManiscalco2023}. In fact, a collisional model was recently implemented to study the steady state of an XXZ spin system, and even reach a (periodic) non-equilibrium steady state \cite{mi2023stable}. 

Here we look into the practical feasibility of using collisional models to simulate dissipative boundary-driven quantum systems. 
In particular, extending the work in \cite{mi2023stable}, we focus on the number of collisions and depths of the circuit required to reach the steady state and characterize how close such a state is to the ideal scenario of infinitely short and weak collisions.  
We find that a nonequilibrium steady state can be reached with circuits of depth within reach of current digital quantum computers, although with still some limitations on the size of the systems that can be studied, and how well they can simulate the dissipative dynamics of a local master equation in Gorini-Kossakowski-Sudharshan-Lindblad form \cite{GoriniSudarshan1976, Lindblad1976}.  
For these setups, we also study the robustness of the results versus errors in the implementation of the collisions.  
To further investigate the performance of the collisional models and their eventual implementation on a digital quantum computer, we also consider a model which leads to strong spin current rectification \cite{LandiSchaller2022}.  

The paper is structured as follows: in Sec.~\ref{intro_CM}, we introduce the collisional model we use in our simulations and how it is implemented. 
Sec.~\ref{XXZ_rectifier} presents the spin chain Hamiltonians considered and their properties.  Sec.~\ref{results} shows the results for the uniform spin chain both in parameter regimes for which a long spin chain would be in the ballistic or diffusive regimes. In Sec.~\ref{rectification} we extend our analysis to the XXZ Hamiltonian with a non-uniform magnetic field which shows strong rectification. In Sec.~\ref{sec:hardware}, we discuss more in detail hardware implementations. Finally, we draw our conclusions in Sec.~\ref{sec:conclusions}. 
\begin{figure}
    \centering
    \includegraphics[width=\columnwidth]{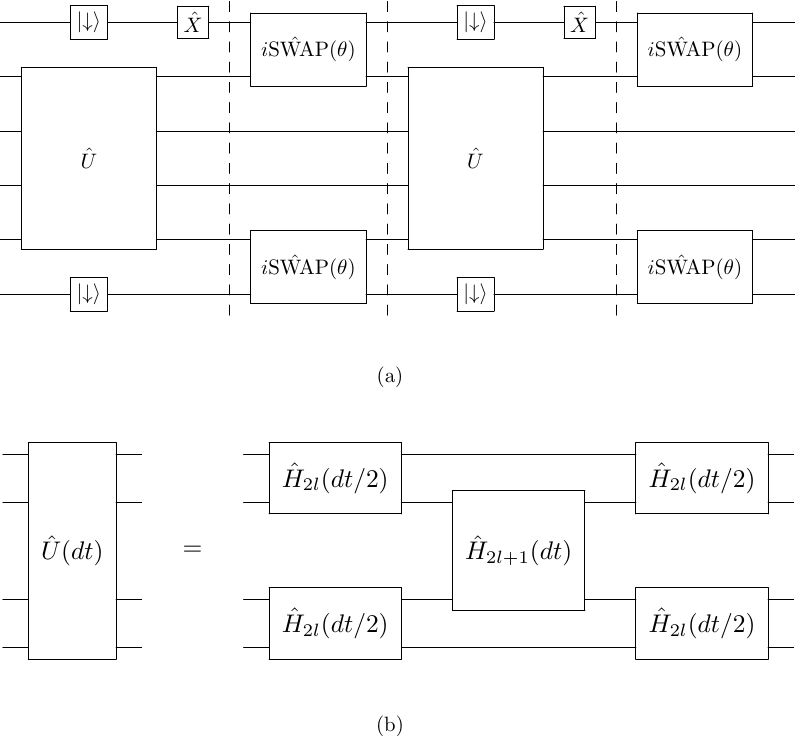}
 \caption{(a) Diagram of a collisional model on a system of 4 sites (qubits $q_1$ to $q_4$) and 2 baths (qubits $q_0$ and $q_5$). Here we show two collisions, each composed of two layers divided by a vertical dashed line. In the first layer we have, in parallel, the unitary evolution of the 4 system qubits, and the measurement and resetting of the ancilla ones. The second layer shows the partial swaps between the two bath qubits and the qubits at the extremities of the system. (b) Trotter step of the unitary evolution as presented in Eq.~(\ref{eq:suzuki-trotter2}). }
\label{fig:CM_circuit}
\end{figure}

\section{Collisional model} \label{intro_CM}  
The dynamics of a quantum system weakly coupled to a memory-less bath can be described by a master equation in the Gorini-Kossakowski-Sudarshan-Lindblad (GKSL) form \cite{GoriniSudarshan1976,Lindblad1976}. Such master equations can be approximated by a set of repeated interactions, or collisions, between the system and a suitably prepared environment. 
For instance, for each bath the system is in touch with, one can prepare many copies of ancilla qubits, and then sequentially, let the system interact with one of these qubits, per bath, at a time. 
In the following, we refer to these short interactions as collisions. 

This approach requires the use of a large number of ancilla qubits. However, since all ancillas are equally initialized, one could also let the system interact with just one ancilla and reset the latter after an interaction time $\tau$. This strategy, that we employ in the rest of the paper, thus allows to significantly reduce the total number of qubits needed to simulate a dissipative process.

In the following sections, we consider a one-dimensional spin chain coupled at the extremities to two baths. More specifically, we consider a spin chain with $L$ spins, to which we add at each extremity another spin to make the two ancillae spins. We thus have a system of $L+2$ spins where spins $0$ and $L+1$ are the ancillae spins, while spins $1$ to $L$ are the system spins. 
The initial state of the system $\hat{\rho}_S$ plus two environment ancillae, $\hat{\rho}_{E_R}$ and $\hat{\rho}_{E_L}$ respectively for the right and left one, is $\hat{\rho}_{E_L} \otimes \hat{\rho}_S \otimes \hat{\rho}_{E_R}$, i.e., an uncorrelated state between the three parts. Furthermore, each portion is prepared in a pure state. 

The collisional model is thus achieved by the following two steps: (i) the system and the ancillae interact for a time $\tau$ and (ii) the system evolves through unitary evolution determined by its Hamiltonian (decoupled from the ancillae) while, at the same time, the ancillae are reset to their initial state. We note that the actions at step (ii) can be done concurrently because the operators commute. 
A schematic description of the implementation of one iteration of the collisional model for a system of four spins is given in Fig.~\ref{fig:CM_circuit}(a). 

The overall evolution is thus determined by two unitary evolutions, $\hat{U}_S=\exp(-\im \hat{H}_S \tau)$ which evolves the system due to its Hamiltonian $\hat{H}_S$ for a time $\tau$, and $\hat{U}_I=\exp(-\im \hat{H}_I \tau)$ which evolves the ancillae and the system for a time $\tau$ via the interaction Hamiltonian $\hat{H}_I$. 
In the following, the interaction operator $\hat{U}_I$ between the system and the ancillae takes the form of a partial $\hat{i\rm{SWAP}}$, which is a native gate for superconducting circuits \cite{YanOliver2018,GoogleAIQuantumMartinis2020}. The choice of the partial $\hat{i\rm{SWAP}}$ is because we aim to simulate the scenario in which the bath is trying to impose its magnetization. This could be done by using a $\hat{\rm{SWAP}}$ gate, but we noticed that a partial $\hat{i\rm{SWAP}}$ allows reaching a steady state with a much smaller number of collisions. 
As a reminder, the $\hat{\rm{SWAP}} $ operation $\hat{S}$ can be expressed in terms of the Pauli operators as 
\begin{equation}
    \hat{S}=\frac{1}{2}(\hat{\mathbb{1}}\otimes \hat{\mathbb{1}} +\hat{Z}\otimes \hat{Z}+\hat{X}\otimes \hat{X}+\hat{Y}\otimes \hat{Y}),
\label{eq:swap_paulis}
\end{equation}
and swaps the states of two qubits
\begin{equation}
    \hat{S}\ket{\psi}\ket{\phi}=\ket{\phi}\ket{\psi}.
\label{eq:swap}
\end{equation}
The partial $\hat{i \rm{SWAP}}$ is then
\begin{equation}
    \hat{U}_I=e^{-\im \hat{S}\theta}=\hat{\mathbb{1}}\cos(\theta)-\im \hat{S}\sin(\theta).
\end{equation} 
A small value of $\theta$ implies a small swap between the system and the ancilla, stemming from a short duration of the interaction between them. 
In App.~\ref{ME_derivation}, we show that one can derive a GKSL-type master equation for the dynamics of the reduced density matrix of the system by considering infinitesimal collisions, and we show that $\theta$ with $\tau$ are related by \cite{BreuerPetruccione2007, LandiLandi} 
\begin{equation}
    \theta={\rm asin}\left(\sqrt{1-e^{-\gamma \tau}}\right). 
\label{eq:theta_main} 
\end{equation}

\section{The boundary-driven XXZ spin chains}\label{XXZ_rectifier}

In order to obtain general enough results, we use a prototypical Hamiltonian for the spin chain, i.e. the XXZ model with, in some scenarios, an external inhomogeneous magnetic field. 
The Hamiltonian is given by 
\begin{align}
    \hat{H} =&\sum_{l=1}^{L-1} [J(\hat{\sigma}_l^{x}\hat{\sigma}_{l+1}^{x} + \hat{\sigma}_l^{y}\hat{\sigma}_{l+1}^{y}) +\Delta \hat{\sigma}_l^{z}\hat{\sigma}_{l+1}^{z} ] +\sum_{l=1}^{L} h_l \hat{\sigma}_l^{z},
\label{eq:ham1}
\end{align}
where $\hat{\sigma}_l^{\alpha=x,y,z}$ the Pauli matrices, $J$ the magnitude of the tunneling between nearest sites, $\Delta$ the magnitude of the coupling between the spins, $h_l$ the site-dependent magnetic field in the $z$ direction. For simplicity here we only consider the magnetic field to be site dependent. The ratio $\Delta/J$ is often referred to as the anisotropy parameter. This is a prototypical model to study quantum transport \cite{BertiniZnidaric2021, LandiSchaller2022}. We work in units such that $\hbar=1$ and $J=1$.    

During the collisions, the first and last spins are coupled to two single-spin baths at sites $l=0$ and $l=L+1$, prepared (and reset) in the up and down states i.e. $\hat{\rho}_{E,L}=\ket{\uparrow}\bra{\uparrow}$, $\hat{\rho}_{E,R}=\ket{\downarrow}\bra{\downarrow}$. We choose this set-up for the bath because (i) it can lead to stronger currents, (ii) stronger effects of the interaction \cite{RossiniZnidaric, BenentiRossini2009} and (iii) these are states which are very easy to prepare in a quantum circuit (for example compared to mixtures or other correlated states). 
In the limit of strong and instantaneous collisions, the dissipative evolution of the reduced density matrix of the system is described by a GKSL master equation \cite{GoriniSudarshan1976,Lindblad1976}
\begin{equation}
    \frac{d\hat{\rho}_S}{dt}=-\im[\hat{H},\hat{\rho}_S]+\sum_{l=1,L}\textit{D}_l (\hat{\rho}_S) ,
\label{eq:lindblad}
\end{equation}
where $D_l$ are the dissipators acting on the first and last sites. The general expression for the dissipators is given by 
\begin{align}
    \textit{D}_l (\hat{\rho}_S) =\gamma&\left[\lambda_l \left(\hat{\sigma}_l^{+}\hat{\rho}_S \hat{\sigma}_l^{-}-\frac{1}{2}\{\hat{\sigma}_l^{-}\hat{\sigma}_l^{+},\hat{\rho}_S \} \right) \right.\nonumber\\    
    &\left.+(1-\lambda_l) \left(\hat{\sigma}_l^{-}\hat{\rho}_S\hat{\sigma}_l^{+}-\frac{1}{2}\{\hat{\sigma}_l^{+}\hat{\sigma}_l^{-},\hat{\rho}_S \} \right)\right],
\label{eq:dissipator}
\end{align}
where $\gamma$ is the dissipation rate due to the baths,  while the driving imposed by the baths is set by $\lambda_l$  and $\hat{\sigma}_l^{\pm}=(\hat{\sigma}_l^{x} \pm \im \hat{\sigma}_l^{y})/2$ are the raising and lowering operators. 
When $\lambda_l=0$ ($\lambda_l=1$) on one side of the chain, the dissipator alone tends to set the spin to be pointing down $\ket{\downarrow}_l\bra{\downarrow}$ (up $\ket{\uparrow}_l\bra{\uparrow}$). For $\lambda_l=0.5$ instead, the dissipator drives the spin is acting on, towards an equal mixture of up and down $(\ket{\downarrow}_l\bra{\downarrow} +\ket{\uparrow}_l\bra{\uparrow})/2$. 

Once the system reaches the steady state, i.e. $d\hat{\rho}_S/dt=0$, there will be a steady spin current flowing through the system, which can be computed as
\begin{equation}
    \mathcal{J}=\mathrm{tr}(\hat{j}_l \hat{\rho}_S),
\label{eq:current}
\end{equation}
where $\hat{j}_l$ is the current operator for the $l$-th bond 
\begin{equation}
    \hat{j}_l=2 J (\hat{\sigma}_l^{x}\hat{\sigma}_{l+1}^{y}-\hat{\sigma}_l^{y}\hat{\sigma}_{l+1}^{x}). 
\label{eq:bond_current}
\end{equation}

In the case in which $\lambda_1 = 1$  and $\lambda_L = 0$ (or vice versa) and the absence of magnetic field $h_l=0$, one can observe different transport regimes in the system, notably a ballistic regime for $0\le \Delta<1$ and diffusive for $\Delta>1$ \cite{LandiSchaller2022,BertiniZnidaric2021}. 
In the presence of a large magnetic field that points in one direction for half a chain, and in the opposite direction for the other half, the emerging steady state can have extremely different spin current magnitudes depending on whether $\lambda_1=0$ and $\lambda_L=1$ or vice versa. In other words, the system is a strong spin current rectifier \cite{LenarcicProsen2015, LeePoletti2020, LeePoletti2021}.

For the unitary evolution of the system, we implement a second-order Suzuki-Trotter decomposition \cite{HatanoSuzuki2005,BerrySanders2007}.  The Hamiltonian Eq.~(\ref{eq:ham1}) can, in fact, be written as a sum of nearest neighbor terms on site $l$ and $l+1$ alone, which we refer to as $\hat{H}_l$, and thus $\hat{H}=\sum_l \hat{H}_l$. 
The unitary evolution on the system alone $U(dt)$ for a short time $dt$ can thus be approximated by 
\begin{align}
    \hat{U}(dt)\!\!=& \!\!\left(\prod_{l} e^{-\im \hat{H}_{2l} \frac{dt}{2}} \right) \!\!\! \left(\prod_{l} e^{-\im \hat{H}_{2l+1} dt}\right) \!\!\! \left(\prod_{l} e^{-\im \hat{H}_{2l} \frac{dt}{2}} \right),   
\label{eq:suzuki-trotter2} 
\end{align} 
where we have grouped together the Hamiltonian terms acting on odd and even bonds.  
For the overall unitary evolution $\hat{U}(\tau)$ one thus needs to repeat $\hat{U}(dt)$ a $\tau/dt$ number of times, noting that $\prod_{l} e^{-\im \hat{H}_{2l+1} \frac{dt}{2}}$ at the end of a $\hat{U}(dt)$ can be merged in a single operation with the one at the beginning of the next $dt$.

\section{Results} \label{results}  
In Sec.~\ref{sec:XXZ} we show results for the XXZ model in Eq.~(\ref{eq:ham1}) without external fields ($h_l=0$), in Sec.~\ref{sec:xxznoise} we consider the effect of errors in the implementation of the unitary evolution and of the dissipation, and in  Sec.~\ref{rectification} we show the results for the XXZ model in Eq.~(\ref{eq:ham1}) with non-zero $h_l$ such that the steady state would have strong rectification.

\subsection{XXZ model}\label{sec:XXZ} 
\begin{figure}[ht!]
\includegraphics[width=\columnwidth]{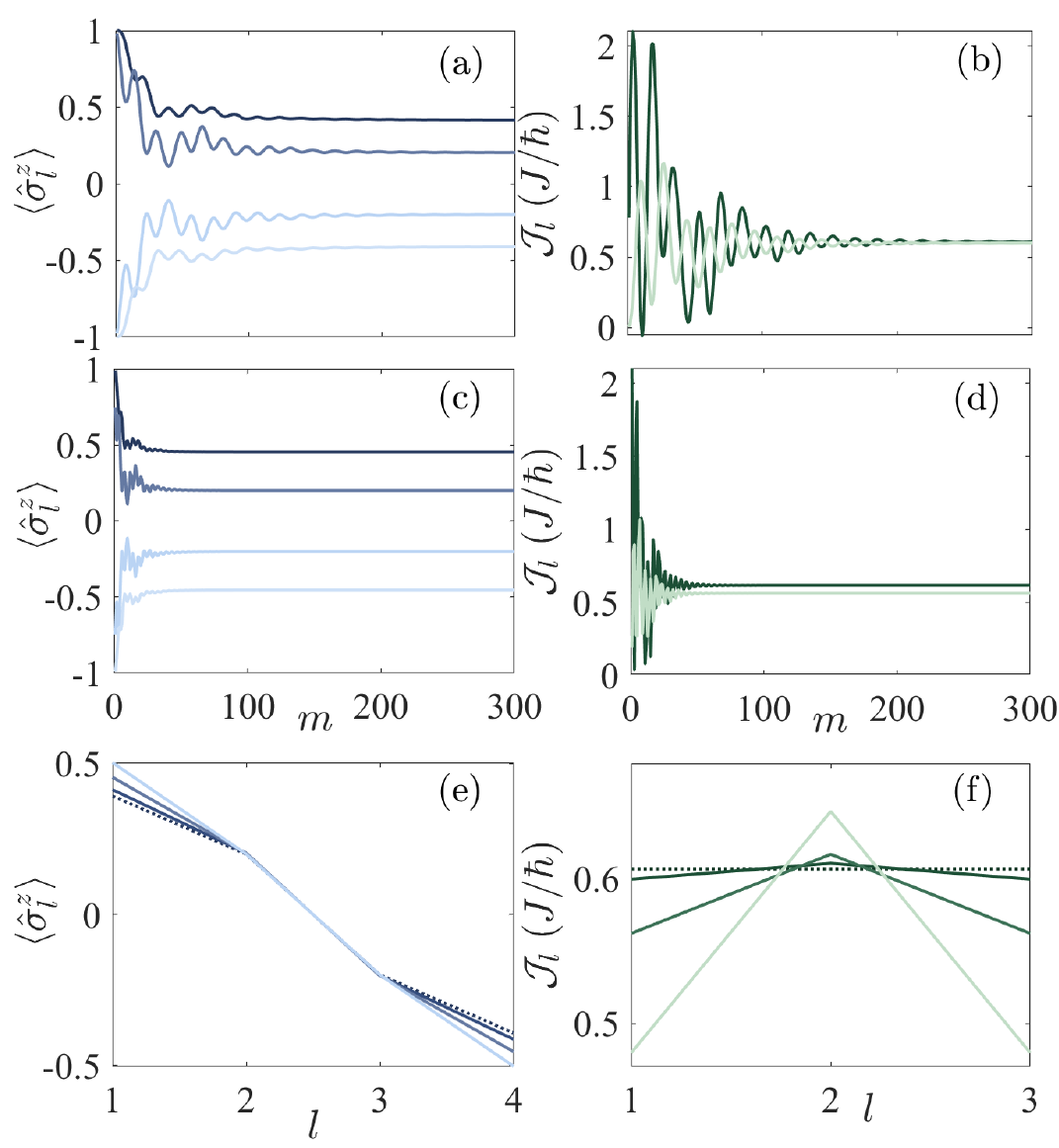}
\caption{Magnetization profile and spin current for a spin chain with $L=4$ spins, coupled to a left and right bath in the up and down states respectively. The parameters of the system and interaction Hamiltonians and the Suzuki-Trotter evolution are $h_z=0$, $\Delta=1.5$, $\gamma=1$, $\tau=0.05$ ($\tau=0.2$), $dt=0.01$, $ \theta=0.222 (\theta=0.439)$ for the first (second) row. The third row shows a summary of the magnetization profiles and steady-state currents for different values of $\tau = 0.05, 0.2, 0.4 $ from darker to lighter. As $\tau \rightarrow 0$ the result should approach the solution to the GKSL master equation. } \label{fig:fig1}
\end{figure}

\begin{figure}[ht!]
\includegraphics[width=\columnwidth]{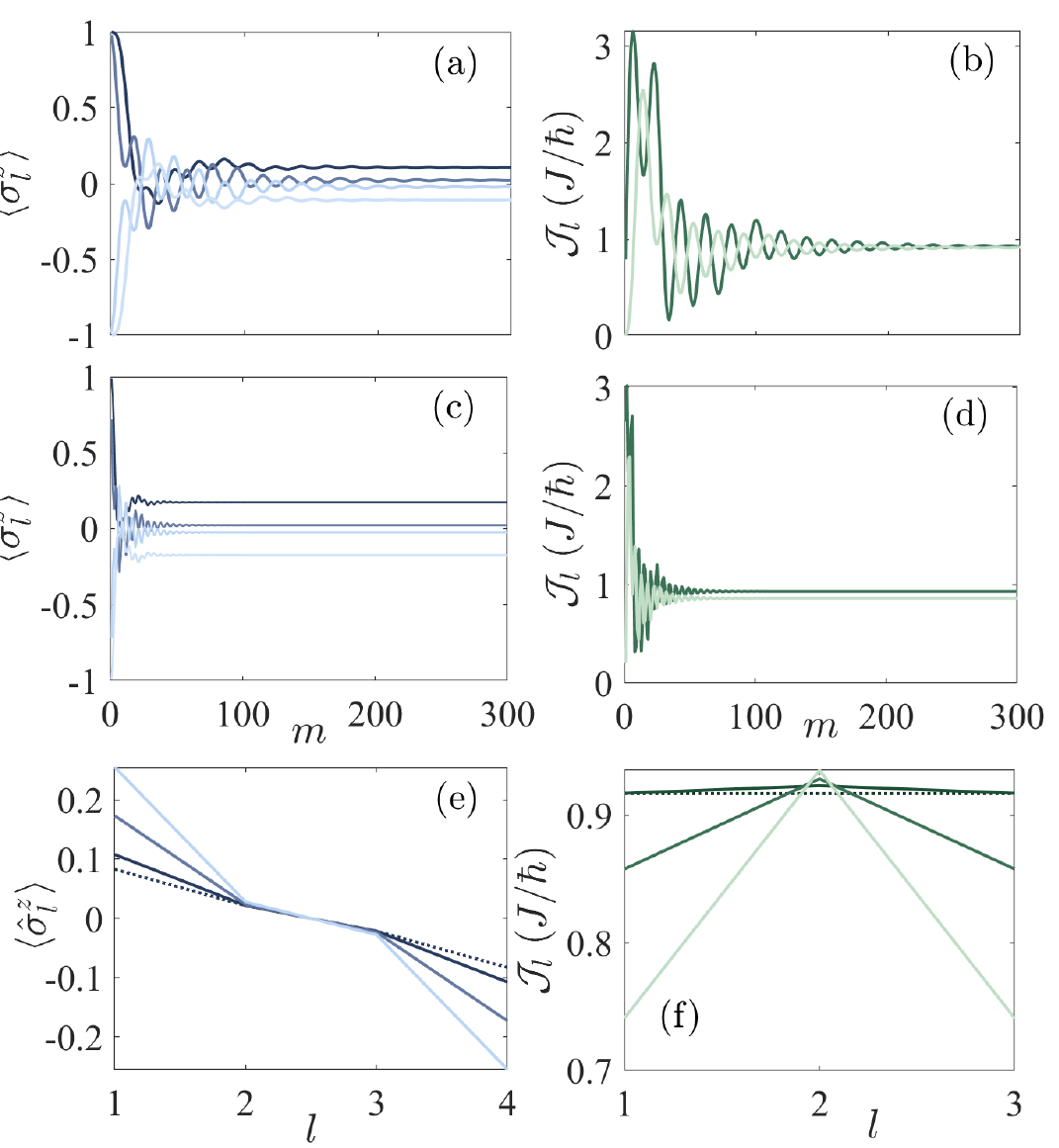}
\caption{Magnetization profile and spin current for a spin chain with $L=4$ spins, coupled to a left and right bath in the up and down states respectively. The parameters of the system and interaction Hamiltonians and the Suzuki-Trotter evolution are: $h_z=0$, $\Delta=0.5$, $\gamma=1$, $\tau=0.05$ ($\tau=0.2$), $dt=0.01$, $\bm{ \theta=0.222 (\theta=0.439)}$ for the first (second) row. The third row shows a summary of the magnetization profiles and steady-state currents for different values of $\tau = 0.05, 0.2, 0.4 $ from darker to lighter. } \label{fig:fig2}
\end{figure}
For concreteness, we now consider a system with $L=4$, $\lambda_1=1$, $\lambda_L=0$ and $\Delta=1.5J$ or $\Delta=0.5J$. The first case is represented in Fig.~\ref{fig:fig1} while the second in Fig.~\ref{fig:fig2}.  
In all the computations below we use the state $\hat{\rho}_S=\ket{\uparrow\uparrow\downarrow\downarrow}\bra{\uparrow\uparrow\downarrow\downarrow}$ as our initial condition. 
In both figures, in panels (a) and (b) we consider a short interaction time (and Suzuki-Trotter step) $\tau=0.05$ while in panels (c) and (d) we show the results for $\tau=0.2$ \footnote{We have chosen values of $\tau$ which can be small enough to give results similar to the master equation, but also not too small that would require a too large number of collisions.}. In panels (a) and (c) we show the local magnetization at each site, while in panels (b) and (d) we show the current in each of the three bonds in the system. 
Panel (e) and (f) depict, respectively, the magnetization versus the site and the current versus the bonds, in the steady state for different values of the collision time $\tau$. Each $\tau$ is represented by a continuous line from darker to lighter color for smaller to larger $\tau$. The steady-state values expected by an exact reproduction of the master equation \ref{eq:lindblad} are depicted by the dotted line. 

In both Fig.~\ref{fig:fig1}(a-d) and Fig.~\ref{fig:fig2}(a-d) we observe that larger values of $\tau$ result in the system reaching a steady state with a smaller number of collisions $m$, shown clearly both by the magnetization and current plots. However, for larger values of $\tau$ one observes that the current is not the same in the different bonds, something we would expect in a steady state. We relate this to the discretized evolution with different operation acting in different intervals. This also implies that the system has actually reached a periodic steady state whose temporal oscillations are more pronounced if $\tau$ is larger. 
From panels (e,f) we also observe that the magnetization and current values approach more accurately the expected value from the master equation only for very small $\tau$, which implies a much larger number of collisions. The difference is more pronounced for the more weakly interacting case $\Delta=0.5$ (for which a long chain would be ballistic), Fig.~\ref{fig:fig2}, which we associate with the larger current present in the system.

\subsection{XXZ model with errors}\label{sec:xxznoise} 

\begin{figure} 
\includegraphics[width=\columnwidth]{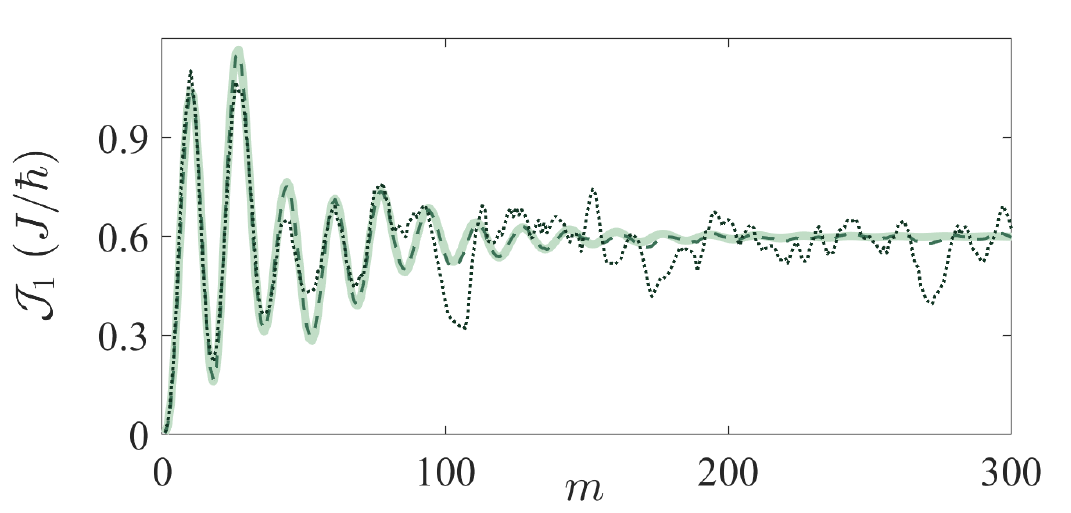}
\caption{This plot shows the evolution of the spin current on the first bond as a function of the number of collisions, for different magnitudes of noise. We consider the parameters of the system and interaction Hamiltonians, as well as the reinitialization of the baths, to be affected by Gaussian white noise with mean $\mu=0$ and different values of standard deviation $\sigma$. From the plot we can see that up to $\sigma=10^{-2}$, the evolution can be robust to noise and the current shows a good overlap with the noiseless scenario. All lines have been computed starting from a set of common parameters: $L=4$, $\tau=0.05$ ($\theta=0.22$), $dt=0.01$, $h_z=0$, $\Delta=1.5$ and $\gamma=1$.} \label{fig:fig3}
\end{figure}

An implementation of the collisional model on a quantum computer cannot currently occur without errors. Here we are considering that at each time step $dt$ we add a random number drawn from an unbiased normal distribution with standard deviation $\sigma$ to the terms $J$ and $\Delta$ in the Hamiltonian. We also add or subtract the modulus of a number drawn from the same distribution to the coefficients $\lambda_l$ which are meant to reset the environment spins at sites $0$ and $L+1$. In particular we add it for $\lambda_L$ and subtract it for $\lambda_1$ such that $0\le \lambda_l \le 1$. Furthermore, we also add a random number drawn from an unbiased distribution with variance $\sigma$ to the collision parameter $\theta$. 
The resulting current versus the number of collisions for different values of the standard deviation $\sigma$ are shown in Fig.~\ref{fig:fig3}. Here we observe that one can obtain quantitatively accurate results for $\sigma= 10^{-2}, 10^{-3}$, while it is difficult to observe even the emergence of a steady state for $\sigma = 10^{-1}$.

\subsection{XXZ model plus non-uniform field} \label{rectification} 
\begin{figure} 
\includegraphics[width=\columnwidth]{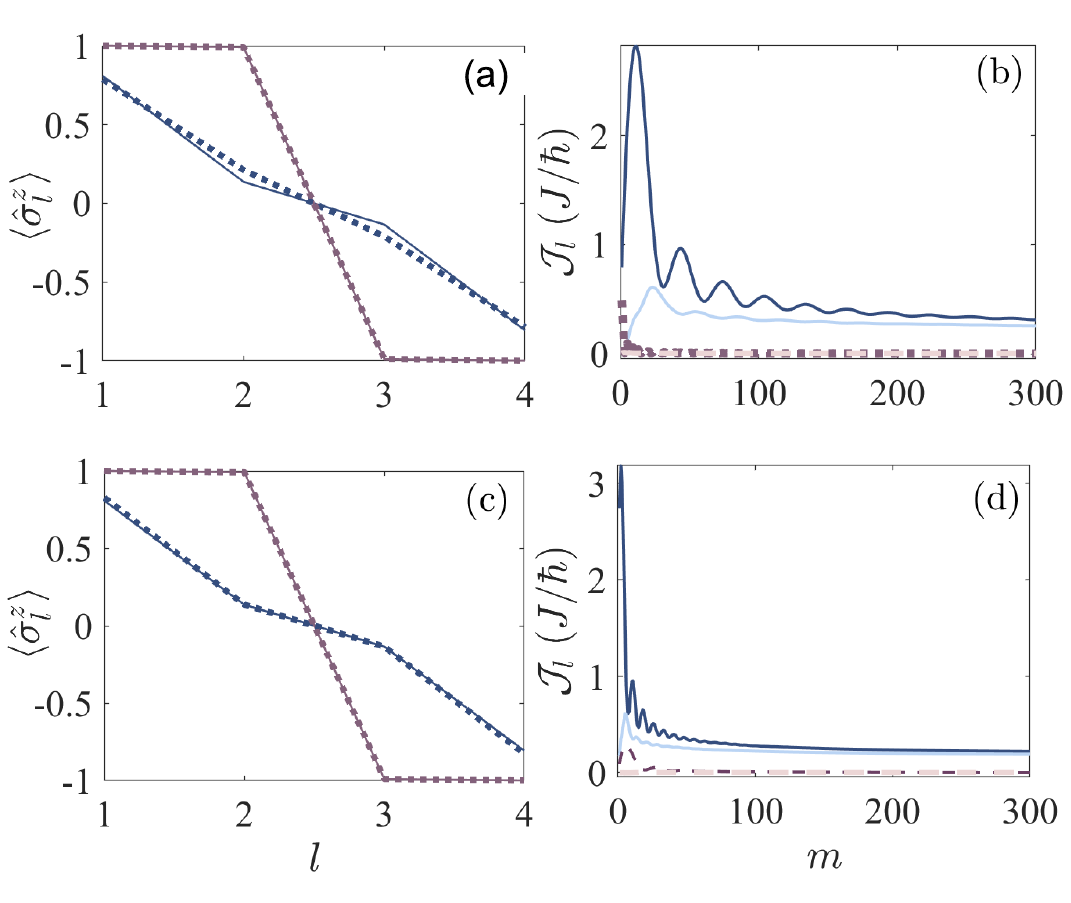}
\caption{This plot shows the cumulative forward and reverse currents and the magnetization profiles for $\tau=0.05$ (first row) and $\tau=0.2$ (second row). The parameters of the system and interaction Hamiltonians and the Suzuki-Trotter evolution are: $L=4$, $h_z=4$, $\Delta=4$, $\gamma=1$, $\tau=0.05$ ($\tau=0.2$), $dt=0.01$, $\theta=0.222\; (\theta=0.439)$ for the first (second) row.} \label{fig:fig4}
\end{figure}

Here we consider the case of a system for which $h_l$ is non-zero in a way that it leads to strong rectification. This can occur when $h_l= h$ for the first half of the spins, and $h_l=-h$ for the other half. In this case, for $\Delta$ and $h$ large and appropriately tuned \cite{LeePoletti2021}, one can observe very little current for   $\lambda_1=1$ and $\lambda_L=0$, but a much larger current for the opposite dissipatively boundary driving $\lambda_1=0$ and $\lambda_L=1$. We refer to the latter as forward bias and the former to reverse bias. We can then define the rectification coefficient as the ratio between the forward and reverse currents 
\begin{equation}
R=-\frac{\textit{$I_f$}}{\textit{$I_r$}}
\end{equation}
where $R=0$ or $\infty$ for a perfect diode, and $R=1$ if there is no rectification. 
In all the computations below we use the state $\hat{\rho}_S=\ket{\uparrow\uparrow\downarrow\downarrow}\bra{\uparrow\uparrow\downarrow\downarrow}$ as our initial condition for the forward bias, and we changed the sign of the local field to -$h$ for the reverse bias scenario.     
In Fig.~\ref{fig:fig4}(b,d) we observe the value of the current at different bonds versus the number of collisions for the forward bias (continuous lines) and reverse bias (dashed lines), while different bonds are differentiated by different shades of the same curve, from lighter to darker for the first to the third bond. In panel (b) we show the result for a smaller $\tau$, while panel (d) for a larger value of $\tau$. Similarly to the results shown in Fig.~\ref{fig:fig1} and \ref{fig:fig2}, for larger $\tau$ one observes a faster convergence towards the steady state. 
When analyzing the difference between the forward and reverse bias currents we observe two striking results: (i) the current in forward bias is indeed significantly larger than in reverse bias and (ii) the reverse bias current converges in a smaller number of steps compared to the forward one. 
The latter point is particularly important because the relaxation gap, i.e. the rate of decay of the slowest eigenvalue of the master equation (\ref{eq:lindblad}), is smaller for the reverse bias. However, the quantitative difference in the current is negligible on the scales set by the forward current. 
Regarding the different magnitude of the currents in the forward and reverse bias, if we consider the smallest value of the current between the different bonds in the forward bias, and the largest current in the reverse bias. We find $R= 186.4$ for $\tau=0.05$ and $R=1219.2$ for $\tau=0.2$.   
In Fig.~\ref{fig:fig4}(a,c) we observe that in reverse bias the steady state magnetization from the collisional model and that from the master equation (\ref{eq:lindblad}), dotted lines, match very well for small $\tau$, panel (a), and larger $\tau$, panel (c). Interestingly, for larger $\tau$ we observe a good match of the magnetization with the one from the master equation, panel (c), and not so well for the smaller $\tau$, panel (a). This is due to the fact that for small $\tau$ the forward bias case has not reached a steady state by the $300-$th collision.   
We conclude this section by stating that for this model with a clear rectification effect, very few collisions are sufficient to tell that there is a significant difference in the current from one bias to the other.

\section{Hardware implementation}\label{sec:hardware} 
We now go back to Fig.~\ref{fig:CM_circuit} to discuss an implementation on a digital quantum computer. 
At a larger scale, each collision can be implemented in two steps. For the first step, one resets the baths and applies the unitary on the system, while in the next step, the collision between the bath qubits and the qubits at the extremities of the system takes place. However, the unitary evolution of the system is composed of smaller blocks, each consisting of an evolution for a time $dt$ of the bond Hamiltonian $\hat{H}_l$. 
One needs to consider how both this two-qubit evolution and the $\hat{i\rm{SWAP}}$ gates are implemented, and this depends significantly on the native gates available. 
For instance, the $\hat{i \rm{SWAP}}$ gate is native on some platforms \cite{YanOliver2018,GoogleAIQuantumMartinis2020,mi2023stable}, and hence this operation can be done with depth one. Interestingly, if we consider the Hamiltonian in Eq.~(\ref{eq:ham1}) with $\Delta=J$, then also each unitary step can be done with a depth of one on these machines. On a machine like IBM Manila, though, the $\hat{i \rm{SWAP}}$ gate is not native and one would need a depth of about 10 to produce both the $\hat{i\rm{SWAP}}$ and the unitary evolution step for a time $dt$ for two qubits, see App.~\ref{app:2spins}.        

Another important consideration when implementing this circuit on a quantum computer, is the duration of each operation, and compare this with the coherence time of the system. 
For instance, if we consider the IBM Manila quantum processor, which has 5 aligned qubits, and the decomposition to native gates of the unitary evolution depicted in Figure \ref{fig:CM_circuit_2}, there are 9 layers of single-qubit gates and 3 layers of two-qubit gates. Considering that on the IBM Manila QPU \cite{Manila} single-qubit operations take $ 35.5$ \unit{\nano\second} (except $\hat{R}_Z$ that comes at zero cost) and that two-qubit operations take $576$ \unit{\nano\second}, the building block of unitary evolution described in Figure \ref{fig:CM_circuit_2} takes 1.9 \unit{\micro\second}. 
At the same time, the algorithm is also resetting the bath qubits and flipping one of them. For the resetting we first perform a measurement and then flip the spin. Since the measurement and resetting takes, in IBM Manila, at least $5.3$ \unit{\micro\second} \footnote{This time is a median time.}.
These timings need to be compared with the coherence times, for this setup the median coherence times are $T_1=188.77$ \unit{\micro\second} and $T_2= 67.3$ \unit{\micro\second} , and from here we can deduce what is the limit of the circuit depth.

Considering a different platform, such as IonQ Harmony \cite{harmony}, we would be dealing with a system with 11 physical qubits with all-to-all connectivity. In this setup, the readout time is 100 \unit{\micro\second} and the reset time is 25 \unit{\micro\second}. These times can be faster than the one and two-qubit gates which are the building blocks of the unitary evolution and the $\hat{i\rm{SWAP}}$s, respectively $10$ \unit{\micro\second} and $210$ \unit{\micro\second}, while the coherence times are $T_1>10^{7}$ \unit{\micro\second} and $T_2= 2\times 10^5$ \unit{\micro\second}. This would allow a sizeable amount of collisions to take place making approaching the steady state possible.

We have mentioned in the early parts of the paper that one can reset the ancilla qubits so that it is no longer needed to keep one ancilla qubit per collision. In a setup in which intermediate measurements are particularly time-consuming or can lead to unwanted effects, it is however possible to also consider what we refer to as a hybrid approach. 
In this case one would use a certain number of available ancilla qubits to perform the partial swaps, one ancilla qubit at the time. Then, one can reset them all together once they have each gone through the partial swap with the system qubits.

\section{Conclusions}\label{sec:conclusions}
We have analyzed the performance of using a collisional model to study a dissipative boundary-driven system, with a steady state with non-zero current, on a quantum computer. This is possible in setups that allow partial measurements on the system without affecting the rest of the setup. This can be done, for instance, on some trapped ions realizations of digital quantum computers \cite{GaeblerPino2021, decross2022qubitreuse}, and also on superconducting chips \cite{rudinger2021characterizing}, provided the time to execute a measurement is short enough. 

From our computations, we observe that it is possible to reach steady states with a limited number of collisions, which could be possible to observe in near-term simulators or digital computers. However, to observe quantitative matches between results expected from a master equation one would need a significant number of collisions. 
For more complex scenarios, like systems that have strong rectification, the forward bias seems much harder to observe as it requires even more collisions to reach a steady regime, however, one can observe a significantly different dynamics already after a few collisions from the forward to the reverse bias. This could be used as an indicator of a potential strong rectifier. 

Another important matter is the scalability of this approach. We have thus also tested the effectiveness of this method for systems with only two spins, see Appendix \ref{app:2spins}. We have observed that the system converges to a steady state significantly faster than for 4 spins, after about $25$ collisions which correspond to about $50$ implementations of macro gates like the unitaries on the system qubits and the swaps, for the case with $\Delta=1.5$, $h_z=0$ and $\tau=0.2$. These macro gates would be implemented by a series of native gates, see for instance App.~\ref{app:2spins}, however an ion-trap setup, because of the speed of measurement and of the large coherence times, could be able to approach the steady state. 
Future work could consider different implementations of master equations on a quantum computer, such as variational approaches \cite{EndoYuan2020}, or using finite size baths \cite{XuPoletti2022, XuPoletti2023}.

\section{Acknowledgments} 
D.P. acknowledges support from the National Research Foundation, Singapore under its QEP2.0 program (NRF2021-QEP2-02-P03). 
D.P. also acknowledges fruitful discussions with K. P. Putti.

\appendix
\section{Derivation of the master equation in a collisional model with a Markovian environment} \label{ME_derivation} 
\subsection{Derivation of the GKSL master equation in a collisional model}

In this section, we show that in the limit of instantaneous and infinitesimal collisions, and non-Markovian interactions, the evolution of the system can be reduced to that of one governed by a GKSL master equation \cite{CattaneoManiscalco2022,CusumanoCusumano2022,BreuerPetruccione2007}. 
Let us consider a system described by a density matrix $\rho_S$ and a set of $K$ identical and uncorrelated ancillae described by $\hat{\rho}_A=\bigotimes_{m=1}^{M}\hat{\rho}_m $. If one assumes the system to be initially uncorrelated to the environment, their joint density matrix at time step $0$ is 
\begin{equation}
    \hat{\rho}_{SA}(0)=\hat{\rho}_S(0) \otimes \hat{\rho}_A
\end{equation}
where $\rho_S$ represents the density matrix of the system. 
The system then evolves under a unitary operator $\hat{U}_S$ for a certain time $\tau$, after which it interacts with the $m$-th ancilla via $\hat{U}_{SA_m}$. Then it evolves again with $\hat{U}_S$ and interacts with the $(m+1)$-th ancilla via the unitary operator $\hat{U}_{SA_{m+1}}$, and so forth. After the collision with the $m$-th collision, where each interaction lasts a time $\tau$, the state of the marginal density matrix of the system is 
\begin{equation}
    \hat{\rho}_S(m\tau)=\tr_{A_m}\left\{\hat{U}_{SA_m} \left[\tilde{\hat{\rho}}_S((m-1) \tau)\otimes \hat{\rho}_{A_m} \right] \hat{U}_{SA_m}^{\dagger}\right\}      
\label{eq:rho_t}
\end{equation}
where $\tilde{\hat{\rho}}_S((m-1) \tau)=\hat{U}_S(\tau)\hat{\rho}_S((m-1) \tau)\hat{U}_S^\dagger(\tau)$. 
\begin{equation}
    \hat{U}_{SA_m}=e^{-\im g\hat{H}_{SA_m}\tau}=e^{-\im\theta \hat{S}}
\label{eq:USA}
\end{equation}
with $\hat{H}_S$ the Hamiltonian that describes the free evolution of the spin chain and $\hat{H}_I$ refers to the interaction Hamiltonian between the system and baths. Here $\tau$ is the unitary evolution time as well as the interaction time and g is the interaction strength. 
If, as it is in our numerical simulations, the system interacts with the ancillae through a partial $\hat{i\rm{SWAP}}$, then $\hat{H}_{SA_i}=\hat{S}$ and $\theta=g\tau$.\\
It is easy to see that in order to derive a GKSL-type master equation, one needs to expand Eq.~(\ref{eq:USA}) to second order in $\theta$ 
\begin{equation}
    \hat{U}_{SA}=\hat{\mathbb{1}}-\im\theta \hat{S}-\frac{\theta^{2}}{2}\hat{S}^{2}+\mathcal{O}(\theta^{3})
\label{eq:U_expansion}
\end{equation}
In order to derive the evolution of $\hat{\rho}_S(m \tau)$ in time, we insert Eq.~(\ref{eq:U_expansion}) in (\ref{eq:rho_t}) and obtain
\begin{align}
    \frac{\hat{\rho}_S(m\tau)-\hat{\rho}_S((m-1)\tau)}{\tau}&=\frac{\theta^{2}}{\tau} \tr_{A_m}[\hat{S}(\hat{\rho}_S \otimes \hat{\rho}_{A_i})\hat{S}-\nonumber \\& \frac{1}{2}\{\hat{S}^{2},\hat{\rho}_S((m-1)\tau)\otimes \hat{\rho}_{A_{m}} \}]
\label{eq:CMevolution}
\end{align}
where we applied the stability condition
\begin{equation}
    \tr_{A_i}[\hat{S}(\hat{\rho}_S((m-1)\tau)\otimes \hat{\rho}_{A_m})]=0 
\end{equation}
to remove any correction coming from the system-baths interaction to $\hat{H}_S$. 

Comparing Equation \ref{eq:CMevolution} to the standard GKSL master equation, we can derive the expression of the system-ancilla coupling strength $\gamma$, in the limit of instantaneous collisions and infinite g 
\begin{equation}
    \gamma=   \lim_{g\to\infty,\tau \to 0} g^{2}\tau =\lim_{g\to\infty,\theta \to 0} g \theta
\end{equation} 
Note that, unless one is interested in studying the correlations between the system and all the ancillae it has interacted with, he/she can use a single ancilla but then reset it to its initial value after the collision. 
In this framework, one does not need to rely on the Born and Markov approximation, since in our collisional model the ancillae are initially uncorrelated and are immediately reinitialized after each interaction with the system. Because each collision is a CP operation, the secular approximation is also not needed to guarantee the complete positivity of the dynamical evolution, unlike in typical microscopic derivations of the master equation.

\subsection{Amplitude damping in a collisional model}
In order to compare the results obtained with a collisional model in the limit of instantaneous collisions, to those of a GKSL master equation, one needs an expression linking the time $\tau$ of the unitary evolution and the system-bath coupling strength $\gamma$, to $\theta$ \cite{BreuerPetruccione2007, LandiLandi}. Here we are going to first give an expression of how a qubit changes under the effect of an amplitude damping channel. Then we show the connection between this perspective and that of using a GKSL master equation. Last we show how a partial SWAP can correspond to an amplitude damping channel and thus connect the partial SWAP with the GKSL master equation. 

In a collisional model, the role of an amplitude damping channel is taken by the partial $\hat{i\rm{SWAP}}$ operator defined in Eq.~(\ref{eq:USA}), which models the loss of energy of the system due to its interaction with the baths \cite{NielsenChuang2000}.\\
As every CPTP map admits a Kraus decomposition, i.e. it can be cast to the operator-sum representation
\begin{equation}
    \hat{\mathcal{E}}(\hat{\rho})=\sum_k \hat{M}_k \hat{\rho} \hat{M}_k^{\dagger} \hspace{0.4cm} \text{with  } \sum_k \hat{M}_k^{\dagger}\hat{M}_k=\hat{\mathbb{1}}
\label{eq:quantum_channel}
\end{equation}
we can consider the Kraus operators of the amplitude damping channel on a 2-level system
\begin{equation}
\hat{M}_0=\begin{pmatrix}
1 & 0 \\
0 & \sqrt{1-\lambda} 
\end{pmatrix},
\hat{M}_1=\begin{pmatrix}
0 & \sqrt{\lambda}  \\
0 & 0
\end{pmatrix}
\label{eq:kraus_AD}
\end{equation}
with $\lambda\in [0,1]$. The effect of $\hat{\mathcal{E}}_{AD}(\rho)$ on a general density matrix $\rho$ is 
\begin{equation}
    \hat{\rho}=\begin{pmatrix}
    p& q\\
    q^{*}&1-p
    \end{pmatrix}\longrightarrow \hat{\rho}'=\begin{pmatrix}
    \lambda+ p(1-\lambda)& q\sqrt{1-\lambda}\\
    q^{*}\sqrt{1-\lambda}&(1-p)(1-\lambda)
    \end{pmatrix}
\label{eq:rho_prime}
\end{equation}
where $p$ and $q$ refer to the population and coherence.
Starting from a generic form of the GKSL master equation, we can now derive an expression of $\lambda$ as a function of $\gamma$
\begin{equation}
    \frac{d\hat{\rho}}{d\tau}=\gamma[\hat{\sigma}_{+}\rho\hat{\sigma}_{-}-\frac{1}{2}\{\hat{\sigma}_{-}\hat{\sigma}_{+},\hat{\rho}\}]
\label{eq:dissipator2}    
\end{equation}
from which we get
\begin{equation}
    \frac{dp}{d\tau}=\gamma(1-p), \hspace{0.5cm} \frac{dq}{d\tau}=-\gamma \frac{q}{2}
\end{equation}
with solutions
\begin{equation}
    p(t)=p_0e^{-\gamma \tau}+(1-e^{-\gamma \tau})
\end{equation}
\begin{equation}
    q(t)=q_0e^{-\gamma \tau/2}
\end{equation}
and if we compare them to $\hat{\rho}'$ in Eq.~(\ref{eq:rho_prime}), we finally get
\begin{equation}
    \lambda=1-e^{-\gamma \tau}
\label{eq:lambda}
\end{equation}
which gives the probability of losing a photon to the environment. Then, the action of the amplitude damping channel is stronger when $\lambda$ is closer to 1, and its effect is that of diagonalizing $\hat{\rho}$ i.e. destroying its coherences, and bringing the system close to the $\ket{0}\bra{0}$ state.\\
The last step is that of introducing parameter $\theta$ of the collisional model into this setting. This is readily done by considering the effect of the partial $\hat{i\rm{SWAP}}$ channel on the joint density matrix of the system $\hat{\rho}$, defined in Eq.~(\ref{eq:rho_prime}), and an ancilla initialized in the down state, i.e. 
\begin{equation}
    \hat{\rho}_{SA}=\hat{\rho}_S \otimes |\downarrow\rangle \langle \downarrow|_A. 
\end{equation} 
The unitary evolution on the system and ancilla $\hat{U}_{SA}$ hence becomes 
\begin{align}
    \hat{U}_{SA}=e^{-\im\theta \hat{S}}&=\hat{\mathbb{1}}\cos(\theta)-\im \hat{S}\sin(\theta)=\\& \begin{pmatrix}
        1& 0& 0& 0\\
        0& \cos(\theta)& -\im\sin(\theta)& 0 \\
        0 &-\im\sin(\theta)& \cos(\theta) &0 \\
        0& 0& 0& 1
    \end{pmatrix}
\label{eq:U_swap}
\end{align}
with
\begin{equation}
    \hat{S}=\frac{1}{2}(\hat{\mathbb{1}}\otimes \hat{\mathbb{1}} +\hat{Z}\otimes \hat{Z}+\hat{X}\otimes \hat{X}+\hat{Y}\otimes \hat{Y}). 
\label{eq:swap_paulis2}
\end{equation}
Then we have that the updated density matrix of the system plus ancilla becomes 
\begin{equation}
    \hat{\rho}_{SA}'=\hat{U}_{SA}\hat{\rho}_{SA}\hat{U}_{SA}^{\dagger}
\end{equation}
and tracing out the environment results in the updated density matrix of the system 
\begin{equation}
    \hat{\rho}_{S}'=\begin{pmatrix}
        p + (1 - p)\sin^{2}(\theta) & q\cos(\theta)\\
        q^{*}\cos(\theta) & (1-p)\cos^{2}(\theta)\\
    \end{pmatrix}
\end{equation}
which, compared to Eq.~(\ref{eq:rho_prime}) and using Eq.~(\ref{eq:lambda}), finally gives
\begin{equation}
    \theta={\rm asin}\left(\sqrt{1-e^{-\gamma \tau}}\right). 
\label{eq:theta}
\end{equation}
 Hence, in the limit of instantaneous collisions, one can use Eq.~(\ref{eq:theta}) to reproduce the results of the standard GKSL master equation, in the setting of a collisional model.

\section{Two-spin system}\label{app:2spins}

\begin{figure}[h!]
\includegraphics[width=\columnwidth]{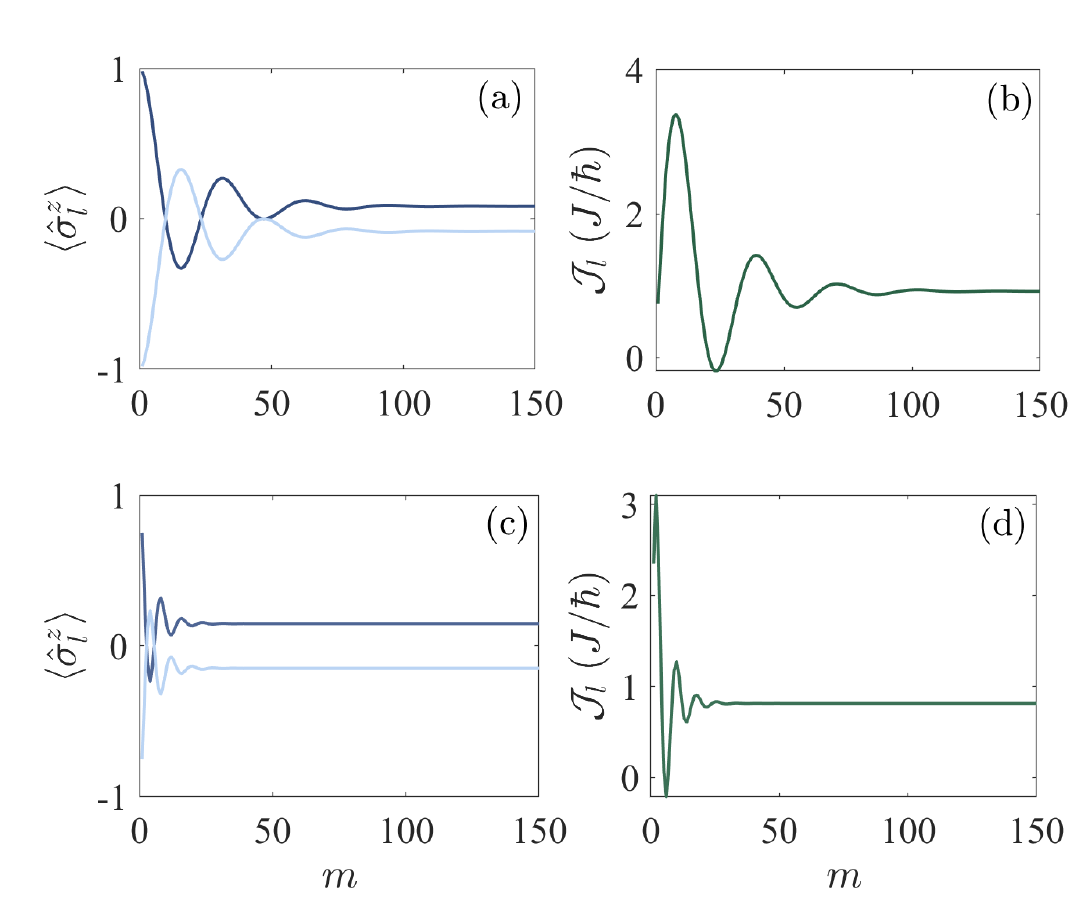}
\caption{(a,c) Magnetization profile and (b,d) spin current for a spin chain with $L=2$ spins, coupled to a left and right bath in the up and down states respectively. The parameters of the system and interaction Hamiltonians and the Suzuki-Trotter evolution are $h_z=0$, $\Delta=1.5$, $\gamma=1$, $dt=0.01$, $\tau=0.05$, i.e. $ \theta=0.222 $, panels (a,b), and $\tau=0.2$, i.e. $\theta=0.439$, panels (c,d). } \label{fig:figB1} 
\end{figure}

\begin{figure*}
    \centering
\includegraphics[width=1\linewidth]{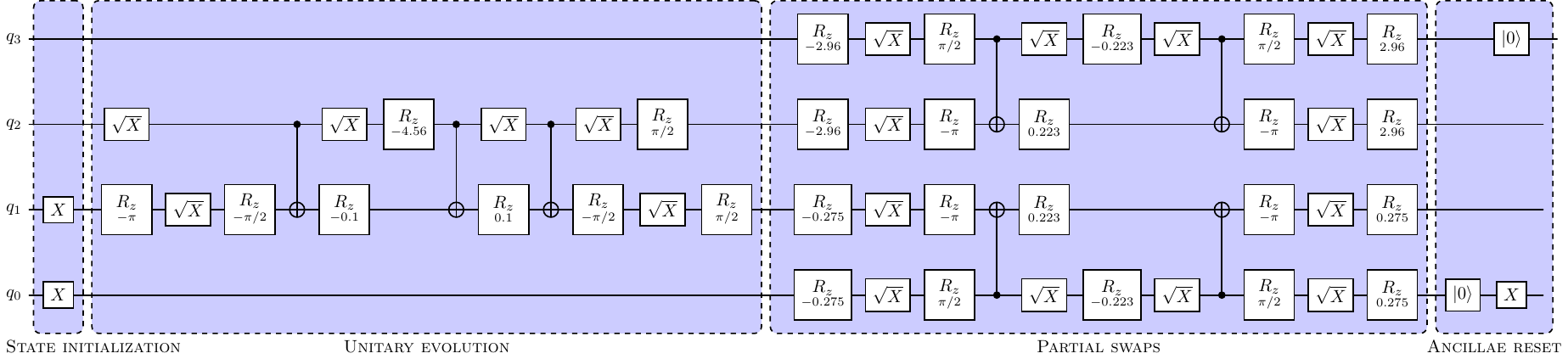}
    \caption{Example of decomposition of a quantum circuit for a collisional model with $m=1$ on two sites and two baths, in terms of gates directly implementable on the quantum processor IBM Manila. For simplicity, here the unitary evolution is computed via second-order Suzuki-Trotter with $dt=\tau$. }
    \label{fig:CM_circuit_2}
\end{figure*}
It is important to check how the collisional model performs differently for different system sizes. Here we analyze the simplest case of a dissipative boundary-driven system with transport, which we take as a two-spin system. In Fig.~\ref{fig:figB1} we observe that for similar parameters as Fig.~\ref{fig:fig1}, the system reaches a steady state in a much smaller number of steps, for example in panels (c) and (d) it takes about 25 collisions. \\
However, implementing each collision requires a circuit depth that depends on the processor used. If we consider, for instance, IBM Manila, both the unitary evolution and the partial swaps are implemented with a depth $\approx 10$, see Fig.~\ref{fig:CM_circuit_2}. Note that here, for clarity of illustration, we picture the ancillae reset and the unitary evolution as happening consecutively but they can, in principle, be performed simultaneously. This implies that to reach the 25 collisions needed to approach the steady state, one would need a circuit depth of about 500. It is thus important to use a processor in which the $\hat{i\rm{SWAP}}$ gate, or any other gate which preserves the total magnetization, is native as this can reduce the depth of the circuit significantly.      
\newpage
\bibliography{collision}
\end{document}